\documentclass[12pt]{article}
\usepackage{amsmath,amsfonts,amssymb}
\usepackage{epsfig,graphicx,bm}
\usepackage{epstopdf}
\begin{document}

\begin{titlepage}

\begin{center}

\vskip 0.4 cm

\begin{center}
{\Large  \bf Restricted $f(R)$ Gravity and Its
 Cosmological Implications }
\end{center}

\vskip 1cm

\vspace{1em} M. Chaichian,$^a$ A. Ghalee,$^b$ J. Kluso\v{n},$^{c,}$
%Markku Oksanen,$^a$ and Anca Tureanu
%$^{a,}$
\footnote{Email addresses:
masud.chaichian@helsinki.fi (M. Chaichian), ghalee@ut.ac.ir (A. Ghalee), klu@physics.muni.cz (J.
Kluso\v{n}), }\\
\vspace{1em}$^a$\textit{Department of Physics, University of Helsinki,
P.O. Box 64,\\ FI-00014 Helsinki, Finland}\\
\vspace{.3em} $^b$\textit{Department of Physics, Tafresh University, Tafresh, Iran}\\
\vspace{.3em} $^c$\textit{Department of Theoretical Physics and
Astrophysics, Faculty of Science,\\
Masaryk University, Kotl\'a\v{r}sk\'a 2, 611 37, Brno, Czech Republic}

\vskip 0.8cm

\end{center}

\begin{abstract}
We investigate the $f(R)$ theory of gravity with broken diffeomorphism
due to the change of the coefficient in front of the total
divergence term in the ($3+1$)-decomposition of the scalar curvature. We
perform the canonical analysis of this theory and show that its
consistent, i.e. with no unphysical degrees of freedom, form is equivalent to the low-energy limit of the non-projectable
$f(R)$ Ho\v{r}ava-Lifshitz theory of gravity. We also analyze
its cosmological solutions and show that
 the de Sitter solution can be obtained
 also  in  the case of this  broken symmetry. The consequences of
 the proposed theory on the asymptotic solutions of a few  specific models in the cosmological context are also presented.
\end{abstract}

\end{titlepage}

\bigskip

\newpage

\def\bn{\mathbf{n}}
\newcommand{\bC}{\mathbf{C}}
\newcommand{\bD}{\mathbf{D}}
\def\hf{\hat{f}}
\def\tK{\tilde{K}}
\def\tmG{\tilde{\mG}}
\def\mC{\mathcal{C}}
\def\bk{\mathbf{k}}
\def\tpi{\tilde{\pi}}
\def\tp{\tilde{p}}
\def\tr{\mathrm{tr}\, }
\def\tmH{\tilde{\mH}}
\def\tf{\tilde{f}}
\def\tY{\mathcal{Y}}
\def\nn{\nonumber \\}
\def\bI{\mathbf{I}}
\def\tmV{\tilde{\mV}}
\def\e{\mathrm{e}}
\def\bE{\mathbf{E}}
\def\bX{\mathbf{X}}
\def\bY{\mathbf{Y}}
\def\bR{\bar{R}}
\def\hN{\hat{N}}
\def\hK{\hat{K}}
\def\hnabla{\hat{\nabla}}
\def\hc{\hat{c}}
\def\mH{\mathcal{H}}
\def \Gi{\left(G^{-1}\right)}
\def\hZ{\hat{Z}}
\def\bz{\mathbf{z}}
\def\bK{\mathbf{K}}
\def\iD{\left(D^{-1}\right)}
\def\tmJ{\tilde{\mathcal{J}}}
\def\tr{\mathrm{Tr}}
\def\mJ{\mathcal{J}}
\def\partt{\partial_t}
\def\parts{\partial_\sigma}
\def\bG{\mathbf{G}}
\def\str{\mathrm{Str}}
\def\Pf{\mathrm{Pf}}
\def\bM{\mathbf{M}}
\def\tA{\tilde{A}}
\newcommand{\mW}{\mathcal{W}}
\def\bx{\mathbf{x}}
\def\by{\mathbf{y}}
\def \mD{\mathcal{D}}
\newcommand{\tZ}{\tilde{Z}}
\newcommand{\tW}{\tilde{W}}
\newcommand{\tmD}{\tilde{\mathcal{D}}}
\newcommand{\tN}{\tilde{N}}
\newcommand{\hC}{\hat{C}}
\newcommand{\hg}{g}
\newcommand{\hX}{\hat{X}}
\newcommand{\bQ}{\mathbf{Q}}
\newcommand{\hd}{\hat{d}}
\newcommand{\tX}{\tilde{X}}
\newcommand{\calg}{\mathcal{G}}
\newcommand{\calgi}{\left(\calg^{-1}\right)}
\newcommand{\hsigma}{\hat{\sigma}}
\newcommand{\hx}{\hat{x}}
\newcommand{\tchi}{\tilde{\chi}}
\newcommand{\mA}{\mathcal{A}}
\newcommand{\ha}{\hat{a}}
\newcommand{\tB}{\tilde{B}}
\newcommand{\hrho}{\hat{\rho}}
\newcommand{\hh}{\hat{h}}
\newcommand{\homega}{\hat{\omega}}
\newcommand{\mK}{\mathcal{K}}
\newcommand{\hmK}{\hat{\mK}}
\newcommand{\hA}{\hat{A}}
\newcommand{\mF}{\mathcal{F}}
\newcommand{\hmF}{\hat{\mF}}
\newcommand{\hQ}{\hat{Q}}
\newcommand{\mU}{\mathcal{U}}
\newcommand{\hPhi}{\hat{\Phi}}
\newcommand{\hPi}{\hat{\Pi}}
\newcommand{\hD}{\hat{D}}
\newcommand{\hb}{\hat{b}}
\def\I{\mathbf{I}}
\def\tW{\tilde{W}}
\newcommand{\tD}{\tilde{D}}
\newcommand{\mG}{\mathcal{G}}
\def\IT{\I_{\Phi,\Phi',T}}
\def \cit{\IT^{\dag}}
\newcommand{\hk}{\hat{k}}
\def \cdt{\overline{\tilde{D}T}}
\def \dt{\tilde{D}T}
\def\bra #1{\left<#1\right|}
\def\ket #1{\left|#1\right>}
\def\mV{\mathcal{V}}
\def\Xn #1{X^{(#1)}}
\newcommand{\Xni}[2] {X^{(#1)#2}}
\newcommand{\bAn}[1] {\mathbf{A}^{(#1)}}
\def \bAi{\left(\mathbf{A}^{-1}\right)}
\newcommand{\bAni}[1]
{\left(\mathbf{A}_{(#1)}^{-1}\right)}
\def \bA{\mathbf{A}}
\newcommand{\bT}{\mathbf{T}}
\def\bmR{\bar{\mR}}
\newcommand{\mL}{\mathcal{L}}
\newcommand{\mbQ}{\mathbf{Q}}
\def\mat{\tilde{\mathbf{a}}}
\def\mtF{\tilde{\mathcal{F}}}
\def \tZ{\tilde{Z}}
\def\mtC{\tilde{C}}
\def \tY{\tilde{Y}}
\def\pb #1{\left\{#1\right\}}
\newcommand{\E}[3]{E_{(#1)#2}^{ \quad #3}}
\newcommand{\p}[1]{p_{(#1)}}
\newcommand{\hEn}[3]{\hat{E}_{(#1)#2}^{ \quad #3}}
\def\mbPhi{\mathbf{\Phi}}
\def\tg{\tilde{g}}
\newcommand{\phys}{\mathrm{phys}}
\def\tmC{\tilde{\mC}}

%%%%%%%%%%%%%%%%%%%%%
%%%%Introduction %%%%%%%%%
%%%%%%%%%%%%%%%%%%%%
\section{Introduction }\label{first}
Recent observation data show that $R^2-$inflation model for the
early Universe is in remarkable agreement with the observations.
\cite{Ade:2015lrj}. Further, the celebrated cosmological constant
problem can be also explained in the context of $f(R)$ theories of
gravity  \footnote{For a review of $f(R)$ theories of gravity, see for
instance \cite{DeFelice:2010aj}.}. It is also possible to find a
formulation of $f(R)$ gravity whose solutions  of equations of
motion capture both the inflation and late time behaviour of the
Universe.

It is important to stress that all these cosmological solutions
depend only on the time, so that they break the manifest
 four-dimensional diffeomorphism of General Relativity. Then one can ask
the question whether the time asymmetry of Friedmann-Robetson-Walker
Universe could be also reproduced by some theory of gravity with
restricted symmetry group.  Indeed, the celebrated
Ho\v{r}ava-Lifshitz (HL) gravity
\cite{Horava:2009uw,Horava:2008ih,Horava:2008jf} which is an
interesting proposal for a renormalizable theory of gravity is based
on the idea of the restricted invariance of the theory when the
theory is invariant under the so-called foliation preserving
diffeomorphism. It turns out that there are two versions of HL
gravity, the projectable theory when the lapse depends only  on the
time  and the non-projectable theory with the lapse depending on the
spatial coordinates too \cite{Horava:2009uw}. It was  subsequently
shown in \cite{Li:2009bg} that the  first versions of
non-projectable Ho\v{r}ava-Lifshitz gravity possesses a pathological
behaviour since the Hamiltonian constraint is a second-class
constraint with itself, which implies that the phase space is
odd-dimensional  (for further discussions, see
\cite{Henneaux:2009zb}). The resolution of this problem was first
suggested in \cite{Blas:2009qj} and further elaborated in
\cite{Blas:2010hb,Blas:2009ck}, where it was argued that in a theory
with broken full diffeomorphism invariance,
 it is natural to include all the  terms which
contain the spatial gradients of lapse and which are invariant under
spatial diffeomorphism too. Then the Hamiltonian analysis of this
theory shows that this Hamiltonian constraint is a second-class
constraint with the primary constraint $\pi_N\approx 0$, where
$\pi_N$ is the momentum conjugate to the  lapse $N$
\cite{Kluson:2010nf,Donnelly:2011df,Mukohyama:2015gia,Chaichian:2015asa}.
On the other hand, the fact that the number of constraints is less
than in General Relativity implies the existence of  an extra scalar
degrees of freedom.

These considerations suggest that the full diffeomorphism invariance
must not be the fundamental symmetry of gravity. In that case it is natural to
consider the possibility whether the $f(R)$ theory of gravity, which  is
not invariant under the full four-dimensional diffeomorphism, can
be consistent with   the  recent cosmological models. There are
certainly several ways how to break full diffeomorphism invariance. The most
natural way is to generalize HL gravity to its $f(R)-$like form as
was done in several papers; see for example
\cite{Kluson:2009rk,Kluson:2010za,
Kluson:2010xx,Kluson:2009xx,Carloni:2010nx,Chaichian:2010yi,Chaichian:2010zn}.
 Recently a new interesting proposal for a theory with the broken full
diffeomorphism invariance was proposed in \cite{Ghalee:2015hwa},
which is based on  the following simple modification
\footnote{Four-divergence  term with coefficient different from $1$
was firstly discussed in \cite{Carloni:2010nx,Chaichian:2010yi} when
the synthesis of $f(R)$ gravity and HL was proposed.}
\begin{equation}\label{prop}
R\rightarrow R_{\Upsilon}\equiv R+(\Upsilon-1)\Xi \ ,
\end{equation}
where $\Upsilon$ is a parameter and $\Xi$ is a four-divergence term
that appears in the decomposition of the Ricci scalar in four
dimensions as \cite{Gourgoulhon:2007ue}
\begin{equation}
R={}^{(3)}R+(K^{ij}K_{ij}-K^2)+\Xi \ .
\end{equation}
Parameter $\Upsilon$ has now meaning as the deformation parameter
that in similar context was introduced in case of $f(R)$ HL gravity
in \cite{Carloni:2010nx,Chaichian:2010yi}
 while its more physical
interpretation will be given in section \ref{fifth} and in
\ref{sixth}. It is important to stress that the last term is a
four-divergence and hence for the Einstein-Hilbert action the
modification (\ref{prop}) does not make sense, since it gives the
boundary contribution that does not affect the equations of motion.
On the other hand, it could have non-trivial consequences in the
case of
 $f(R)$ theories
of gravity as was shown in \cite{Ghalee:2015hwa}.

Due to the interesting features of  this simple idea (\ref{prop}),  we believe  it deserves to be studied further. In
particular we would like to give a more physical justification for it
and see whether the presumptions which were implicitly used in
\cite{Ghalee:2015hwa} have solid physical grounds. In particular, it is
not quite  clear whether the  Newtonian gauge used there could be
implemented in the theory with  the broken diffeomorphism invariance.
Motivated by these facts,  we perform the Hamiltonian analysis of
the restricted $f(R)$ gravity. We show, in agreement with
\cite{Ghalee:2015hwa}, that the full diffeomorphism invariance is broken. On the
other hand, we show that the naive form of restricted $f(R)$ gravity
possesses the same pathological behaviour as the first versions of
non-projectable Ho\v{r}ava-Lifshitz gravity, whose
description was given above. We resolve this problem in a similar
way as in the case of non-projectable HL gravity. More precisely, we
show under which conditions  the restricted $f(R)$ theory could be
considered as a consistent theory from the Hamiltonian point of view.
In the first case we impose an additional constraint on the theory.
However, a  careful Hamiltonian analysis of this version  will show
that now the theory becomes  invariant under the full four-dimensional
diffeomorphism and that it reduces to the standard Einstein-Hilbert
action with a cosmological constant. The second possibility is to
consider an extended version of restricted $f(R)$ gravity when we
include the terms containing spatial gradients of the lapse. However, in this
case we obtain that this theory could be considered as  a low-energy
limit of the non-projectable HL theory or gravity. We perform the
Hamiltonian analysis of this theory, following
\cite{Kluson:2010nf,Donnelly:2011df,Mukohyama:2015gia,Chaichian:2015asa}
and we show that there is an extra scalar degree of freedom, whose
consequences on the physical spectrum around FRW background should
be taken into account.

Having constructed a consistent modification of  $f(R)$ theory of gravity, we
proceed to the analysis of its cosmological solutions and we find that
the properties of these solutions depend on the value of the
parameter $\Upsilon$ that allow us to obtain new  solutions which do not
exist in the ordinary $f(R)$ theory of gravity.

The structure of this paper is as follows. In the next section
 we introduce the original formulation of restricted
$f(R)$ theory of gravity and perform its Hamiltonian analysis. In
section \ref{third} we study a version of this theory with
an additional constraint imposed and we show that this theory  is equivalent to
the ordinary Einstein-Hilbert one. Then in section \ref{fourth} we
propose an extended form of restricted $f(R)$ theory of gravity with
the spatial gradients of lapse included. In section \ref{fifth} we
study some cosmological solutions of such a  theory. Finally,  in
section \ref{sixth}, Discussion,  we outline our results and suggest a possible
extension of the work.

%%%%%%%%%%%%%%%%%%%%%%%%%%%%%%
\section{Restricted $f(R)$ gravity}\label{second}
Let us consider the $4$-dimensional
manifold $\mathcal{M}$ with the
coordinates $x^\mu \ , \mu=0,\dots,3$,
where $x^\mu=(t,\bx) \ ,
\bx=(x^1,x^2,x^3)$. We assume  that
this space-time is endowed with the
metric $\hat{g}_{\mu\nu}(x)$ with
signature $(-,+,+,+)$. Suppose that
$ \mathcal{M}$ can be foliated by a
family of space-like surfaces
$\Sigma_t$ defined by $t=x^0$. In this work, we are interested in the cosmological implications of our model. So, we will use the flat Friedmann-Robertson-Walker Universe that for which $\Sigma_t=\mathbf{R}^{3}$. So, Let
$g_{ij}, i,j=1,2,3$ denote the
metric on $\Sigma_t$ with its inverse
$g^{ij}$, so that $g_{ij}g^{jk}=
\delta_i^k$. We further introduce the operator
$\nabla_i$, which  is the covariant derivative
defined by the metric $g_{ij}$.
% The
% basic vector fields on $\Sigma_t$ are
% $\partial_i$.
 We  introduce  the
future-pointing unit normal vector
$n^\mu$ to the surface $\Sigma_t$. In
ADM variables we have
$n^0=\sqrt{-\hat{g}^{00}},
n^i=-\hat{g}^{0i}/\sqrt{-\hat{g}^{
00}}$. We also define  the lapse
function $N=1/\sqrt{-\hat{g}^{00}}$ and
the shift function
$N^i=-\hat{g}^{0i}/\hat{g}^{00}$. In
terms of these variables we write the
components of the metric
$\hat{g}_{\mu\nu}$ as
\begin{eqnarray}
\hat{g}_{00}=-N^2+N_i g^{ij}N_j \ ,
\quad \hat{g}_{0i}=N_i \ , \quad
\hat{g}_{ij}=g_{ij} \ ,
\nonumber \\
\hat{g}^{00}=-\frac{1}{N^2} \ , \quad
\hat{g}^{0i}=\frac{N^i}{N^2} \ , \quad
\hat{g}^{ij}=g^{ij}-\frac{N^i N^j}{N^2}
\ .
\nonumber \\
\end{eqnarray}
Then it is easy to see that
\begin{equation}
\sqrt{-\det \hat{g}}=N\sqrt{\det g} \ .
\end{equation}
Further we define the extrinsic
derivative
\begin{equation}
K_{ij}=\frac{1}{2N}
(\partial_t g_{ij}-\nabla_i N_j-
\nabla_j N_i) \ .
\end{equation}
It is well known that the components of
the Riemann tensor can be written in
terms of ADM variables \footnote{For
a review and an extensive list of
references, see
\cite{Gourgoulhon:2007ue}.}.
 For
example, in the case of Riemann curvature
we have
\begin{eqnarray}\label{R}
R&=&K^{ij}K_{ij}-K^2+{}^{(3)}R+\frac{2}{\sqrt{-\hat{g}}}
\partial_\mu(\sqrt{-\hat{g}}n^\mu K)-
\frac{2}{\sqrt{g}N}\partial_i
(\sqrt{g}g^{ij}\partial_j N)\nonumber \\
&=&K^{ij}K_{ij}-K^2+{}^{(3)}R+\Xi \ . \nonumber \\
\end{eqnarray}
The restricted $f(R)$ gravity is based  on the idea that we modify
$R$ in the following way
\begin{equation}
R\rightarrow R+(\Upsilon-1)\Xi \ ,
\end{equation}
where $\Upsilon$ is a constant. Then we consider
the action in the form
\begin{equation}\label{actionNOJI}
S_{f(R)}= \frac{1}{\kappa^2} \int dt d^3\bx \sqrt{g}N f
(R+(\Upsilon-1)\Xi) \ .
\end{equation}
Note that, since $\Sigma_t=\mathbf{R}^{3}$ has no boundary, we have not considered any boundary term in the action.\footnote{Generally, this action should be supplemented with the boundary term
in order to make variation principle well defined
\cite{Gibbons:1976ue}. When one wants to study the isolated objects, e.g. the black holes, such terms are needed when asymptotically flat boundary conditions on $\Sigma_t$ are imposed. In case of standard $f(R)$ theory of gravity
such a boundary term has the form \cite{Guarnizo:2010xr,Dyer:2008hb}
$2\oint_{\partial \Sigma}d^3y \epsilon \sqrt{|h|}f'(R)K$ with
$f'(R)=\frac{df}{dR}$, where $\partial \Sigma$ is the boundary of
the manifold, $h$ is the determinant of the induced metric, $K$ is
the trace of the extrinsic curvature of the boundary $\partial
\Sigma$ and $\epsilon$ is equal to $1$ if $\partial \Sigma$ is
time-like and $-1$ if $\partial \Sigma$ is space-like. Finally
coordinates $y^\alpha$ label the boundary $\partial \Sigma$. We
propose that in case of the restricted $f(R)$ gravity the
corresponding boundary term is simple modification of the boundary
term given above when we replace $R$ with (\ref{repla}). However the
detailed analysis of this boundary contribution is beyond the scope
of this paper and will be performed elsewhere.}\\
Our goal is to perform the Hamiltonian analysis of the action
(\ref{actionNOJI}). We introduce  two auxiliary scalar fields A and
B and rewrite the action in the equivalent form
\begin{equation}\label{SFRa}
S_{f(R)}=\frac{1}{\kappa^2}\int dt d^3\bx \sqrt{g}N (f(A)+B(
R+(\Upsilon-1)\Xi-A)) \ .
\end{equation}
It can be easily seen that the action (\ref{SFRa}) is equivalent to
the action (\ref{actionNOJI}) when we solve the equation of motion
for $B$ and gives $A=R+(\Upsilon-1)\Xi$. Inserting back this result
into (\ref{SFRa}) we obtain (\ref{actionNOJI}). Now using the
explicit form of $\Xi$ we can rewrite the action (\ref{SFRa}) into
the form
\begin{eqnarray}\label{SFR}
S_{f(R)}&=&\frac{1}{\kappa^2}\int dt d^3\bx \left( \sqrt{g}N B(
K_{ij}\mG^{ijkl}K_{kl}+{}^{(3)}R -A)\nonumber \right.
\\
&+& \left. \sqrt{g}N f(A) -2\Upsilon \sqrt{g}N\nabla_n B g^{ij}
K_{ji} +2 \Upsilon
\partial_i B \sqrt{g}g^{ij}
\partial_j N \right) \ ,
\nonumber \\
\end{eqnarray}
where we have introduced the de Witt metric $\mG^{ijkl}$
\begin{equation}
\mG^{ijkl}=\frac{1}{2}(g^{ik}g^{jl}+ g^{il}g^{jk})- g^{ij}g^{kl} \ ,
\end{equation}
and where
\begin{equation}
\nabla_n B=\frac{1}{N}(\partial_t B-N^i\partial_i B) \ .
\end{equation}
From (\ref{SFR}) we find the conjugate momenta
\begin{eqnarray}
\pi^{ij}&=&
\frac{1}{\kappa^2}\sqrt{g}B\mG^{ijkl}K_{kl}-\frac{1}{\kappa^2}
\Upsilon\sqrt{g} \nabla_n Bg^{ij} \ , \quad \pi_N\approx 0 \ , \quad
\pi_i \approx 0 \ ,
\nonumber \\
p_B&=&-\frac{2}{\kappa^2}\Upsilon\sqrt{g}K \ , \quad  p_A\approx 0 \ . \nonumber \\
\end{eqnarray}
%After some manipulation, we obtain an inverse relation between
%$K_{ij}$ and $\pi^{ij}$
%%\begin{equation}
%%\pi\equiv \pi^{ij}g_{ji}=
%%-\frac{2}{\kappa^2}\sqrt{g}B K-\frac{3}{\kappa^2}\Upsilon
%%\sqrt{g}\nabla_n B \ ,
%%\Upsilon\sqrt{g}\nabla_n B=-\frac{\kappa^2}{3}
%%(\pi-\frac{1}{\Upsilon}B p_B)
%%\end{equation}
%%using
%%\begin{equation}
%%g_{ij}\mG^{ijkl}=-2g^{kl}
%%\end{equation}
%%Finally we find
%%\begin{eqnarray}
%%\pi^{ij}= \frac{1}{\kappa^2}\sqrt{g}B \mG^{ijkl}K_{kl}+
%%\frac{1}{3}g^{ij}\pi-\frac{1}{3\Upsilon}g^{ij}Bp_B
%%\nonumber \\
%%\end{eqnarray}
%%and consequently
%\begin{equation}
%K_{ij}=\frac{\kappa^2}{\sqrt{g}B} \mG_{ijkl}\left(\pi^{kl}-
%\frac{1}{3}g^{kl}\pi+\frac{1}{3\Upsilon}g^{kl}Bp_B\right) \ , \quad
%\pi\equiv \pi^{ij}g_{ij} \ .
%\end{equation}
Then it is easy to find
 Hamiltonian density in the form
\begin{eqnarray}
\mH=\partial_t g_{ij}\pi^{ij}+p_B\partial_t B-\mL=
%2N \pi^{ij}K_{ij}+Np_B\nabla_n B
%+ N^i(-2g_{ik} \nabla_j \pi^{kj}+\partial_i Bp_B)-\mL=
%\nonumber \\
%=\frac{1}{\kappa^2} \sqrt{g}N B(
%K_{ij}\mG^{ijkl}K_{kl}+{}^{(3)}R
%+A)-\frac{2N}{\kappa^2}\Upsilon \sqrt{g}\nabla_n BK
%-\nonumber
%\\
%-\frac{1}{\kappa^2}\sqrt{g}N F(A) -
%\frac{1}{\kappa^2}\Upsilon
%\partial_i B \sqrt{g}g^{ij}
%\partial_j N =
%\nonumber \\
%=N[\frac{\kappa^2}{\sqrt{g}B}\pi^{ij}\mG_{ijkl}\pi^{kl}+\frac{\kappa^2}{\sqrt{g}B}
%\frac{1}{3}\pi^2-\frac{\kappa^2}{3\sqrt{g}\Upsilon}p_B \pi-\frac{\kappa^2}{6B\sqrt{g}}
%\pi^2+\frac{\kappa^2 }{3\sqrt{g}\Upsilon} p_B \pi-\frac{\kappa^2}{6\Upsilon^2\sqrt{g}}B p_B^2- \nonumber \\
%-\frac{\kappa^2}{3\sqrt{g}}(\frac{\pi p_B}{\Upsilon}-\frac{1}{\Upsilon^2}Bp_B^2)+
%+\frac{\sqrt{g}}{\kappa^2}B({}^{(3)}R+A)-\frac{1}{\kappa^2}\sqrt{g}F(A)+
%\frac{\Upsilon}{\kappa^2}\partial_i[\sqrt{g}g^{ij}\partial_j B]]+N^i\mH_i=\
%\nonumber \\
%N[\frac{\kappa^2}{\sqrt{g}B}\pi^{ij}g_{ik}g_{jl}-
%\frac{\kappa^2}{3\sqrt{g}}\pi^2-\frac{\kappa^2}{3\sqrt{g}\Upsilon}
%p_B\pi+\nonumber \\
%+\frac{\kappa^2}{6\Upsilon^2\sqrt{g}}Bp_B^2
%+\frac{\sqrt{g}}{\kappa^2}B({}^{(3)}R+A)-\frac{1}{\kappa^2}\sqrt{g}F(A)+
%\frac{2\Upsilon}{\kappa^2}\partial_i[\sqrt{g}g^{ij}\partial_j B]]+\nonumber \\
%+N^i\mH_i
N\mH_T+N^i\mH_i \ ,
\nonumber \\
\end{eqnarray}
where
\begin{eqnarray}
\mH_T&=& \frac{\kappa^2}{\sqrt{g}B}\pi^{ij}g_{ik}g_{jl}\pi^{kl}-
\frac{\kappa^2}{3B\sqrt{g}}\pi^2-\frac{\kappa^2}{3\sqrt{g}\Upsilon}
p_B\pi\nonumber \\
&+&\frac{\kappa^2}{6\Upsilon^2\sqrt{g}}Bp_B^2
-\frac{\sqrt{g}}{\kappa^2}B({}^{(3)}R-A)-\frac{1}{\kappa^2}\sqrt{g}f(A)+
\frac{2\Upsilon}{\kappa^2}\partial_i[\sqrt{g}g^{ij}\partial_j B] \ , \nonumber \\
\mH_i&=&-2g_{ik}\nabla_j \pi^{jk}+p_B\partial_i B \  . \nonumber \\
\end{eqnarray}
Now the requirement of the preservation of the primary constraints
$\pi_N(\bx) \approx 0, \pi_i(\bx)\approx 0 $ and $p_A(\bx)\approx 0$,
implies the following secondary ones
\begin{eqnarray}
\partial_t \pi_N(\bx)&=&\pb{\pi_N(\bx),H}=
 \mH_T(\bx)\approx 0 \ ,
\nonumber \\
\partial_t p_i(\bx)&=&\pb{p_i(\bx),H}=
-\mH_i(\bx)\approx 0 \ , \nonumber \\
\partial_t p_A(\bx)&=&-\frac{1}{\kappa^2}\sqrt{g}B+
\frac{1}{\kappa^2}\sqrt{g}f'(A)\equiv G_A(\bx)\approx 0 \ .
\nonumber \\
\end{eqnarray}
Since $\pb{p_A(\bx),G_A(\by)}=
-\frac{1}{\kappa^2}\sqrt{g}f''(A)\delta(\bx-\by)$, we see that
$(p_A,G_A)$ are the second-class
constraints and hence can be explicitly
solved. In  solving  the first one, we
set $p_A$ strongly zero, while
  solving the second one,
 we find
$f'(A)=B$. Assuming  that
$f'$ is invertible, we can express
$A$ as  a function of $B$ so that
 $A=\Psi(B)$ for some function $\Psi$.
Finally, since
$\pb{\pi^{ij},p_A}=\pb{g_{ij},p_A}=0$, we
see that the Dirac brackets between
the canonical variables coincide with
the Poisson brackets.

Now we proceed to the analysis of the preservation
of all constraints. We begin with the constraint
$\mH_i$,  which  we modify in the following way
\begin{equation}
\tmH_i=\mH_i+p_A\partial_i A=-2g_{ik}\nabla_j\pi^{jk}+
p_A\partial_i A+p_B\partial_i B. \
\end{equation}
It is convenient to define the smeared form of these fields
\begin{equation}
\bT_S(N^i)=\int d^3\bx N^i\tmH_i \ .
\end{equation}
The reason why one considers the smeared forms( i.e. takes the
constraints integrated by multiplying  them  with some smooth
functions)  is to easily deal with the distributions, which is the
usual and rigorous way, since the point-wise constraints are
distributions and  in particular they contain the delta functions
and  their derivatives. In principle, one can perform all the
calculations without their smeared forms, but then more care has to
be taken.
 Then it is easy to show  that $\bT_S(N^i)$ are  the
generators of the spatial diffeomorphism and that they are the
first-class constraints. We rather focus on the analysis of the
Hamiltonian constraint. It is also convenient to introduce as well
the smeared form of the Hamiltonian constraint
\begin{equation}
\bT_T(N)=\int d^3\bx N\mH_T \ .
\end{equation}
Our goal is to perform the calculation of the Poisson bracket
between the smeared forms of the Hamiltonian constraints
$\pb{\bT_T(N),\bT_T(M)}$.
% To do that, we have to use the following
%formulas
%\begin{eqnarray}
%\pb{{}^{(3)}R(\bx),\pi^{ij}(\by)}&=&
%-{}^{(3)}R^{ij}(\bx)\delta(\bx-\by)+ \nabla^i
%\nabla^j \delta(\bx-\by)-g^{ij}
%\nabla_k \nabla^k\delta(\bx-\by) \ ,
%\nonumber \\
%\pb{{}^{(3)}R(\bx),\pi(\by)}&=&-{}^{(3)}R(\bx)
%\delta(\bx-\by)-2\nabla_k\nabla^k\delta(\bx-\by) \ .
%\nonumber \\
%%\nabla^i \nabla^j \mG_{ijkl} \pi^{kl}-
%%g^{ij}\nabla_k\nabla^ k
%%\mG_{ijkl}\pi^{kl}= \nonumber \\
%%=\nabla_k (\nabla_l \pi^{kl})
%%+\frac{1-\lambda}{\lambda D-1}\nabla_i
%%\nabla^i \pi \nonumber \\
%\end{eqnarray}
Then after some careful calculations we find
\begin{eqnarray}\label{pbHam}
\pb{\bT_T(N),\bT_T(M)}&=&\bT_S((N\nabla_jM-M\nabla_jN)g^{ji})
%+\frac{4}{3}(1-\Upsilon) \int d^3\bx
%(M\nabla_i N-N\nabla_iM)\frac{\pi}{B}\nabla^iB+\nonumber \\
%+4(1-\Upsilon)\int d^3\bx(N\nabla_iM-M\nabla_iN)\frac{\pi^{ij}}{B}
%\nabla_jB+\nonumber \\
%+\frac{1-4\Upsilon}{3\Upsilon}\int d^3\bx (M\nabla_i
%N-N\nabla_iM)p_B\nabla^i B+\int d^3\bx (N\nabla_k M-M\nabla_k
%N)(-2\nabla_l \pi^{kl})=\nonumber
%\\
\nonumber \\
&+&\frac{4}{3}(1-\Upsilon) \int d^3\bx
(M\nabla_i N-N\nabla_iM)\frac{\pi}{B}\nabla^iB\nonumber \\
&+&4(1-\Upsilon)\int d^3\bx(N\nabla_iM-M\nabla_iN)\frac{\pi^{ij}}{B}
\nabla_jB\nonumber \\
&+&\frac{\Upsilon-1}{3\Upsilon}\int d^3\bx (N\nabla_iM-M\nabla_i
N)p_B\nabla^i B \ . \nonumber \\
%+\int d^3\bx (N\partial_iM-M\nabla_i N)g^{ij}\mH_j
%\nonumber \\
\end{eqnarray}
The expression on the first line is the standard form of the Poisson
bracket between smeared forms of Hamiltonian constraint for the full
diffeomorphism invariant theory. On the other, the additional terms
in (\ref{pbHam}) which  are proportional to $\Upsilon-1$ do not
vanish on the constraint surface and explicitly show the breaking of
the full diffeomorphism invariance. Moreover, this result suggests
that we have a theory where $\mH_T$ is a second-class constraint,
which is the situation that is known from the analysis of the first
versions of non-projectable Ho\v{r}ava-Lifshitz gravity
\cite{Li:2009bg,Henneaux:2009zb}.  As was argued there, the
existence of \emph{one} second-class constraint implies that the
dimension of the physical phase space is odd, what should not be. It
turns out that there are two possibilities how to resolve this
puzzle. The first one is based on the observation that the right
side of the Poisson bracket (\ref{pbHam})
 vanishes on the constraint surface when $\partial_iB=0$. We
will discuss this case in the next section.
\section{Projectable restricted $f(R)$ gravity}\label{third}
To proceed with the condition $\partial_iB=0$, we introduce the following
decomposition of the scalar field $B$
\begin{equation}
B=\tilde{B}+B_0 \ ,
\end{equation}
where
\begin{equation}
B_0=\frac{1}{\int d^3\bx \sqrt{g}}\int d^3\bx \sqrt{g}B
\end{equation}
 and hence $\int d^3\bx
\sqrt{g}\tilde{B}=0$. Then the condition $\partial_iB=0$
implies $\tilde{B}=K(t)$ for any function $K(t)$. On the other hand, since
$\int d^3\bx \sqrt{g}\tilde{B}=0$, we obtain that $K(t)=0$. In other words,
the condition $\partial_i B=0$ is equivalent to the constraint
\begin{equation}
\Phi_I\equiv \tilde{B}\approx 0 \ .
\end{equation}
Obviously, we have to ensure that this constraint is also preserved
during the time evolution of the system. To do that, we also
decompose the momenta $p_B$ as
\begin{equation}
p_B=\tilde{p}_B+\frac{\sqrt{g}}{\int d^3\bx \sqrt{g}}P_B \ , \quad
P_B=\int d^3\bx p_B \ , \quad \int d^3\bx \tilde{p}_B=0 \ ,
\end{equation}
where we have the following Poisson brackets
\begin{eqnarray}\label{canPB}
& &\pb{B_0,P_B}=1 \ , \quad \pb{\tilde{p}_B(\bx),B_0}=\pb{\tilde{B}(\bx),P_B}=0 \ ,
\quad \nonumber \\
& &\pb{\tilde{B}(\bx),\tilde{p}_B(\by)}=\delta(\bx-\by)-\frac{\sqrt{g}(\by)}{
\int d^3\bz \sqrt{g}(\bz)} \ . \nonumber \\
\end{eqnarray}
Finally we have to analyze the requirement of the preservation of
the constraint $\Phi_I\approx 0$
\begin{eqnarray}
& \partial_t\Phi_I &=\pb{\Phi_I(\bx),H} \nonumber \\
&=&\int d^3\by
N\frac{\kappa^2}{3\sqrt{g}}\left(\tilde{p}_B+\frac{\sqrt{g}} {\int
d^3\bz \sqrt{g}}P_B-\Upsilon\pi\right)\left( \delta(\bx-\by)-
\frac{\sqrt{g}(\by)}{ \int d^3\bz \sqrt{g}(\bz)}\right) \ .
\nonumber \\
%\nonumber \\
%&-&
%%\frac{\kappa^2}{3\sqrt{g}\Upsilon}\pi
%(\delta(\bx-\by)-
%\frac{\sqrt{g}(\by)}{
%\int d^3\bz \sqrt{g}(\bz)}))=0
%\nonumber \\
\end{eqnarray}
In order to preserve  the constraint $\Phi_I\approx 0$, it
is natural to impose the following constraint
\begin{equation}
\Phi_{II}\equiv \frac{1}{\sqrt{g}}\left[\tilde{p}_B+\frac{\sqrt{g}}
{\int d^3\bz \sqrt{g}}P_B -\Upsilon\pi\right]\approx 0 \ .
\end{equation}
Now thanks to the Poisson bracket (\ref{canPB}), we see that there
exists a non-zero Poisson bracket $\pb{\Phi_I(\bx),\Phi_{II}(\by)}\neq 0$,
so that they are the second-class constraints. We recall  that there are still  two
additional second-class constraints $p_A\approx 0, G_A\approx 0$.
Solving these second-class constraints, we obtain the Hamiltonian
constraint $\mH_T$ in the form
\begin{eqnarray}\label{mHTproj}
\mH_T= \frac{\kappa^2}{\sqrt{g}B_0}\left(\pi^{ij}g_{ik}g_{jl}-
\frac{1}{2}\pi^2\right)
-\frac{\sqrt{g}}{\kappa^2}B_0({}^{(3)}R-A)-\frac{1}{\kappa^2}\sqrt{g}f(B_0)
 \ , \nonumber \\
\end{eqnarray}
where we have also solved $G_A$ for $A$ as $A=\Psi(B_0)$. Finally, note
that $\mH_T$ does not depend on $P_B$ and hence $B_0$ is constant
on-shell and we see that (\ref{mHTproj}) corresponds to the
Hamiltonian constraint of General Relativity with a cosmological
constant when $B_0$ is absorbed into the definition of $\kappa$. In
other words, the condition $\tB\approx 0$ implies that the above  considered restricted
$f(R)$ gravity is equivalent to General Relativity. For that reason
we have to consider the second possibility when we abandon the
requirement that $\mH_T$ is a  first-class constraint.
 %%%%%%%%%%%%%%%%%%%%%%%%%
\section{Extended Form of Restricted $f(R)$ gravity}\label{fourth}
In this section we show how to resolve the problem with the naive
existence of the second-class constraint $\mH_T$ in the restricted
$f(R)$ gravity.
 The resolution of this puzzle is based on
  the fact that whenever we accept that some theory is not
invariant under the full diffeomorphism, it is natural to include
all the terms that are compatible with the spatial diffeomorphism in
the definition of the action. In other words, we should consider a
more general version of restricted $f(R)$ gravity that is similar to
the so-named healthy extension of HL gravity
\cite{Blas:2009qj,Blas:2010hb,Blas:2009ck}. In this  section we
consider such a modification of  the restricted $f(R)$ gravity when
we include in  the action additional terms which  are invariant
under spatial diffeomorphism. Following  the discussion performed in
the case of  HL gravity,  we also replace the de Witt metric by a
generalized de Witt metric which  has the form \cite{Horava:2009uw}
\begin{eqnarray}
\tmG^{ijkl}&=&\frac{1}{2}(g^{ik}g^{jl}+g^{il}g^{jk})-\lambda
g^{ij}g^{kl} \ , \quad  \lambda \neq \frac{1}{3} \ , \nonumber \\
\tmG_{ijkl}&=&\frac{1}{2}(g_{ik}g_{jl}+g_{il}g_{jk})-\frac{\lambda}{3\lambda-1}
g_{ij}g_{kl} \ . \nonumber \\
\end{eqnarray}
More importantly, due to the fact that the theory is not invariant
under the full four-dimensional diffeomorphism, it is natural to include
the vector $a_i=\frac{\partial_i N}{N}$ into the definition of the
action. In other words, our extended form of restricted $f(R)$
gravity arises when we perform the replacement
\begin{equation}\label{repla}
R\rightarrow K_{ij}\tmG^{ijkl}K_{kl}+{}^{(3)}R+\Upsilon\Xi +\gamma_1
a_i a^i+\gamma_2 {}^{(3)}R^{ij}a_i a_j \ ,
\end{equation}
where $\gamma_1,\gamma_2$ are the corresponding coupling constants.
Then the action with auxiliary fields $A$ and $B$ has the form
\begin{eqnarray}\label{actextended}
\tilde{S}_{f(R)} &=&\frac{1}{\kappa^2}\int dt d^3\bx \left(
\sqrt{g}N B( K_{ij}\tmG^{ijkl}K_{kl}+{}^{(3)}R+\gamma_1 a_i
a^i+\gamma_2 {}^{(3)}R^{ij}a_i a_j-A)\nonumber \right.
\\
&+&\left. \sqrt{g}N f(A) -2\Upsilon \sqrt{g}N\nabla_n B g^{ij} K_{ji}
+2 \Upsilon
\partial_i B \sqrt{g}g^{ij}
\partial_j N \right) \ .
\nonumber \\
\end{eqnarray}
Now we are ready to proceed to the Hamiltonian analysis of the
theory defined by the action (\ref{actextended}). Following the same
logic as in section (\ref{second}) we find
% From
%(\ref{actextended}) we find
%\begin{eqnarray}
%\pi^{ij}&=&
%\frac{1}{\kappa^2}\sqrt{g}B\tmG^{ijkl}K_{kl}-\frac{1}{\kappa^2}
%\Upsilon\sqrt{g} \nabla_n Bg^{ij} \ , \quad, \pi_N\approx 0 \ ,
%\quad \pi_i \approx 0 \ ,
%\nonumber \\
%p_B&=&-\frac{2}{\kappa^2}\Upsilon\sqrt{g}K \ , \quad  p_A\approx 0 , \  \nonumber \\
%\end{eqnarray}
%%Then after some manipulation we obtain
%%\begin{equation}
%%\pi\equiv \pi^{ij}g_{ji}= \frac{1}{\kappa^2}\sqrt{g}(1-3\lambda)B
%%K-\frac{3}{\kappa^2}\Upsilon \sqrt{g}\nabla_n B \ ,
%%\Upsilon\sqrt{g}\nabla_n B=-\frac{\kappa^2}{3}
%%(\pi+\frac{(1-3\lambda)}{2\Upsilon}B p_B)
%%\end{equation}
%%using
%%\begin{equation}
%%g_{ij}\tmG^{ijkl}=(1-3\lambda)g^{kl}
%%\end{equation}
%%Finally we find
%%\begin{eqnarray}
%%\pi^{ij}= \frac{1}{\kappa^2}\sqrt{g}B \tmG^{ijkl}K_{kl}+
%%\frac{1}{3}g^{ij}\pi+\frac{(1-3\lambda)}{6\Upsilon}g^{ij}Bp_B
%%\nonumber \\
%%\end{eqnarray}
%%and consequently
%so that
%\begin{eqnarray}
%K_{ij}
%%=\frac{\kappa^2}{\sqrt{g}B} \tmG_{ijkl}\left(\pi^{kl}-
%%\frac{1}{3}g^{ij}\pi-\frac{(1-3\lambda)}{6\Upsilon}g^{ij}Bp_B\right)=
%%=\nonumber \\
%=\frac{\kappa^2}{\sqrt{g}B}(\pi_{ij}-\frac{1}{3}g_{ij}\pi)-\frac{\kappa^2}{6\sqrt{g}\Upsilon}g_{ij}
%p_B \ .  \nonumber \\
%\end{eqnarray}
%%using
%%\begin{equation}
%%\tmG_{ijkl}(\pi^{kl}-\frac{1}{3}g^{kl}\pi)=\pi_{ij}-\frac{1}{3}g_{ij}\pi
%%\end{equation}
%Then, in the same way as in the second section, we find
the Hamiltonian
density in the form
\begin{eqnarray}
\mH=\partial_t g_{ij}\pi^{ij}+p_B\partial_t B-\mL=
N\mH_T+N^i\mH_i \ ,  \nonumber \\
% 2N
%\pi^{ij}K_{ij}+Np_B\nabla_n B + N^i(-2g_{ik} \nabla_j
%\pi^{kj}+\partial_i Bp_B)-\mL=
%\nonumber \\
%=\frac{1}{\kappa^2} \sqrt{g}N B( K_{ij}\tmG^{ijkl}K_{kl}-{}^{(3)}R
%-\gamma_1 a_i a^i-\gamma_2 {}^{(3)}R_{ij}g^{ij}a_i a_j
%+A)-\frac{2N}{\kappa^2}\Upsilon \sqrt{g}\nabla_n BK -\nonumber
%\\
%-\frac{1}{\kappa^2}\sqrt{g}N F(A) - \frac{1}{\kappa^2}\Upsilon
%\partial_i B \sqrt{g}g^{ij}
%\partial_j N =
%\nonumber \\
%=N[\frac{\kappa^2}{\sqrt{g}B}(\pi^{ij}-\frac{1}{3}g^{ij}\pi)g_{ik}g_{jl}
%(\pi^{kl}-\frac{1}{3}g^{kl}\pi)+\frac{\kappa^2(1-3\lambda)}{12\sqrt{g}\Upsilon^2}
%p_B^2-\nonumber
%\\
%-\frac{\kappa^2}{3\sqrt{g}\Upsilon}(\pi+\frac{(1-3\lambda)}{2\Upsilon}Bp_B)p_B
%-\frac{\sqrt{g}}{\kappa^2}B({}^{(3)}R +\gamma_1 a_i a^i+\gamma_2
%{}^{(3)}R_{ij}g^{ij}a_i a_j -A)-\nonumber
%\\
%-\frac{1}{\kappa^2}\sqrt{g}F(A)+
%\frac{\Upsilon}{\kappa^2}\partial_i[\sqrt{g}g^{ij}\partial_j
%B]]+N^i\mH_i=
%\nonumber \\
%N[\frac{\kappa^2}{\sqrt{g}B}\pi^{ij}g_{ik}g_{jl}-
%\frac{\kappa^2}{3\sqrt{g}}\pi^2-\frac{\kappa^2}{3\sqrt{g}\Upsilon}
%p_B\pi-\nonumber \\
%-\frac{(1-3\lambda)\kappa^2}{12\Upsilon^2\sqrt{g}}Bp_B^2
%-\frac{\sqrt{g}}{\kappa^2}B({}^{(3)}R +\gamma_1 a_i a^i+\gamma_2
%{}^{(3)}R_{ij}g^{ij}a_i a_j -A)-\frac{1}{\kappa^2}\sqrt{g}F(A)+
%\frac{2\Upsilon}{\kappa^2}\partial_i[\sqrt{g}g^{ij}\partial_j B]]+\nonumber \\
%+N^i\mH_i=N\mH_T+N^i\mH_i \ ,
%\nonumber \\
\end{eqnarray}
where
\begin{eqnarray}
\mH_T&=&\frac{\kappa^2}{\sqrt{g}B}\pi^{ij}g_{ik}g_{jl}\pi^{kl}-
\frac{\kappa^2}{3\sqrt{g}}\pi^2-\frac{\kappa^2}{3\sqrt{g}\Upsilon}
p_B\pi -\frac{(1-3\lambda)\kappa^2}{12\Upsilon^2\sqrt{g}}Bp_B^2
\nonumber \\
 &-&\frac{\sqrt{g}}{\kappa^2}B({}^{(3)}R +\gamma_1 a_i
a^i+\gamma_2 {}^{(3)}R^{ij}a_i a_j
-A)-\frac{1}{\kappa^2}\sqrt{g}f(A)+
\frac{2\Upsilon}{\kappa^2}\partial_i[\sqrt{g}g^{ij}\partial_j B]
 \ , \nonumber \\
\mH_i&=&-2g_{ik}\nabla_j \pi^{jk}+p_B\partial_i B \  . \nonumber \\
\end{eqnarray}
Note that this form of the Hamiltonian constraint is in agreement
with the constraint (up to the potential term and terms containing
$a_i$) found in
\cite{Chaichian:2010zn}.

It is also very important to identify the global constraints which  are
related to the action (\ref{actextended}). In fact, it is easy to
see that there is a primary global constraint \cite{Donnelly:2011df}
\begin{equation}
\Pi_N=\int d^3\bx \pi_N N \ ,
\end{equation}
which  has the following non-zero Poisson brackets with $N$ and $\pi_N$
\begin{equation}\label{PiNai}
\pb{\Pi_N,a_i(\bx)}=0 \ , \quad \pb{\Pi_N,N(\bx)}=-N(\bx) \ , \quad
\pb{\Pi_N,\pi_N(\bx)}=\pi_N(\bx) \ .
\end{equation}
We show below that $\Pi_N$ is a first-class constraint. It turns
out that we have to be careful with the definition of the local and
global constraints. Following the notation used in
\cite{Chaichian:2015asa}, we define a  local constraint as
\begin{equation}
\tpi_N(\bx)=\pi_N(\bx)-\frac{\sqrt{g}(\bx)}{\int d^3\bx N\sqrt{g}}
\Pi_N \ .
\end{equation}
Saying it  differently, we decompose the constraint $\pi_N(\bx)$
into the local and global constraints, $\tpi_N(\bx)$ and $\Pi_N$,
respectively and we denote it {\it symbolically} \footnote{In
\cite{Kuchar:1991xd}  the number of   such  constraints was
symbolically  denoted as $\infty^3-1$.} by "$\infty^3-1$" local
constraints $\tpi_N(\bx)$, as follows from the fact that the
constraint $\tpi_N$ obeys the equation
\begin{equation}
\int d^3\bx N(\bx)\tpi_N(\bx)=0 \ .
\end{equation}
Now together with the global constraint $\Pi_N$,  we have a total number
of "$\infty^3$" constraints, which  is the same as the number of the
original constraints $\pi_N$.

In summary, the Hamiltonian with the primary constraints included
has the form
\begin{equation}
H=\Pi_N+\int d^3\bx (N\mH_T+N^i\tmH_i+v_N\tpi_N+v^i\pi_i+v_A p_A) \
\ .
\end{equation}
Now we have to proceed to the analysis of the preservation of the
primary constraints $\tpi_N\approx 0, \pi_i\approx 0 \ , p_A\approx
0$ and $\Pi_N$. In the case of the constraint $\Pi_N\approx 0$ we obtain
\begin{equation}
\partial_t\Pi_N=\pb{\Pi_N,H}=-\int d^3\bx N\mH_T\equiv -\Pi_T\approx
0 \ ,
\end{equation}
where $\Pi_T=\int d^3\bx N\mH_T\approx 0$ is the global Hamiltonian
constraint \cite{Donnelly:2011df}. The requirement of the
preservation of the constraints $\pi_i\approx 0$ and $p_A\approx 0$
implies the same constraints as in the second section, namely
$\mH_i$ and $G_A$. Finally, the requirement of the preservation of
the constraint $\tpi_N\approx 0$ implies
\begin{eqnarray}
\partial_t \tpi_N(\bx)&=&\pb{\tpi_N(\bx),H}\nonumber \\
&=&-\mH_T-
\frac{2}{\kappa^2}\sqrt{g}[B(\gamma_1  a_i a^i +\gamma_2 R_{ij}a^i
a^j)+\nabla_i[B\gamma_1 a^i+\gamma_2 R_{kl}g^{ki}g^{lj}a_j]]
\equiv -\mC(\bx) \ .  \nonumber \\
\nonumber \\
\end{eqnarray}
However, not all the $\mC(\bx)$ are independent since we have
\begin{eqnarray}
\int d^3\bx N\mC=\Pi_T \ ,
%+\nonumber \\
%+ \int d^3\bx \frac{2N}{\kappa^2}\sqrt{g} [B(\gamma_1  a_i a^i
%+\gamma_2 R_{ij}a^i a^j)+\nabla_i[B\gamma_1 a^i+\gamma_2
%R_{kl}g^{ki}g^{lj}a_j]]
%=\nonumber \\
%=\Pi_0+\frac{2}{\kappa^2} \int d^3\bx N\sqrt{g} B(\gamma_1 a_i a^i
%+\gamma_2 R_{ij}a^i a^j)-\nonumber \\
%-\frac{2}{\kappa^2} \int d^3\bx N\sqrt{g} B a_i(\gamma_1  a^i
%+\gamma_2 R_{ij} a^j)=\Pi_0 \nonumber \\
\end{eqnarray}
where we have ignored the boundary terms. Then we see that it is natural
to introduce "$\infty^3-1$" independent constraints $\tmC(\bx)\approx
0$ defined as
\begin{equation}
\tmC(\bx)= \mC(\bx)-\frac{\sqrt{g}(\bx)}{\int d^3\by N\sqrt{g}}\Pi_T
\ ,
\end{equation}
which  obey the relation
\begin{equation}
\int d^3\bx N(\bx)\tmC(\bx)=0 \ .
\end{equation}
In summary, the total Hamiltonian with all constraints included has
the form
\begin{equation}
H_T=\Pi_T+\Pi_N+\int d^3\bx (N\tmC +N^i\tmH_i+v_N\tpi_N+v^i\pi_i+v_A
p_A+\Gamma^A G_A) \ .
\end{equation}
Now we are ready to study the preservation of all the constraints.
It is easy to show that $\Pi_N,\Pi_T$ are global first class
constraints while  $\tmH_i$ are local first class constraints.
%preserved during the time evolution of the system. In fact, their
%smeared form is the generator of the spatial diffeomorphism.
%Secondly, $\Pi_N\approx 0$ is a  first-class constraint as follows
%from (\ref{PiNai}) together with the fact that
%\begin{eqnarray}
%\pb{\Pi_N,\mC(\bx)}&=&\pb{\Pi_N,\pb{\pi_N(\bx),\int d^3\by N\mH_T}}
%\nonumber \\
%&=&-\pb{\pi_N(\bx),\pb{\int d^3\by N\mH_T,\Pi_N}}- \pb{\int d^3\by
%N\mH_T,\pb{\Pi_N,\pi_N(\bx)}}\nonumber \\
%&=&
%-\pb{\pi_N(\bx),\int d^3\by N\mH_T}-\pb{\int d^3\by N\mH_T,
%\pi_N(\bx)}=0 \ ,  \nonumber \\
%\end{eqnarray}
%using the fact that $\pb{\Pi_N,\mH_T(\bx)}=0$. Then it is easy to
%see that $\pb{\Pi_N,\tmC(\bx)}=0$.
On the other hand $(\tpi_N,p_A,\tmC,G_A)$ and the second class
constraints.
% corresponding matrix for  Poisson bracket
%\begin{equation}
%\pb{\Psi_A(\bx),\Psi_B(\by)}=\triangle_{AB}(\bx,\by) \ ,
%\end{equation}
%with its inverse $\triangle^{AB}$. Further we introduce the constraint
%$\tilde{\Pi}_T$ defined as
%\begin{equation}
%\tilde{\Pi}_T=\Pi_T-\int d^3\bz d^3\bz' \pb{\Pi_T,\Psi_A(\bz)}
%\triangle^{AB}(\bz,\bz')\Psi_B(\bz') ,
%\end{equation}
%which  is a  first-class constraint since
%$\pb{\tilde{\Pi}_T,\Psi_A(\bx)}=0$.
In summary,  we have the following picture of the restricted $f(R)$
gravity. This is a theory which  is invariant under the spatial
diffeomorphism with three local first-class constraints
corresponding to this symmetry. We also have four second-class
constraints. Solving these constraints, we can express $A,p_A$ and
$\tpi_N$ and $N$ as functions of the dynamical variables. The
physical phase space of this theory is spanned by $g_{ij},\pi^{ij}$,
where six of these degrees of freedom can be eliminated by
 gauge fixing of the diffeomorphism constraints
$\tmH_i$. We see that there is a scalar graviton degree of freedom
as in the non-projectable HL gravity with all its consequences on the
physical properties of this theory. Finally, there is also a scalar
degree of freedom $B$ with conjugate momenta $p_B$, as in the ordinary
$f(R)$ theory of gravity.
%%%%%%%%%%%%%%%%%%%%%%%%%%%%%%%%%%
\section{ Cosmological Aspects of the Theory}\label{fifth}
\label{sec:1} In this section we study the restricted $f(R)$ theory
of gravity  in the cosmological context for which the FRW metric is
the preferred coordinate system of the Universe.
 We will consider mechanisms that lead to an accelerated expansion phase.\\
To show the mass scale of modified gravity $M$, it is convenient to
consider the usual $f(R)$-gravity as
\begin{equation}\label{cos1}
\frac{f(R)}{\kappa^2}=\frac{M_{P}^{2}R}{2}+
M^{4}\tf\left(\frac{R}{M^{2}}\right) .
\end{equation}
We also assume that the metric has the  standard  flat FRW form
\begin{equation}\label{cos2}
ds^{2}=-N^{2}(t)dt^{2}+a(t)^{2}dx^{i}dx^{j}\delta_{ij} \ ,
\end{equation}
where $a=a(t)$ is the scale factor. Then the right-hand side of
(\ref{repla}) takes  the following form
\begin{equation}\label{cos3}
R_{\Upsilon}\equiv A+\Upsilon\Xi \ ,
\end{equation}
where
\begin{equation}\label{cos4a}
\Xi=-6\frac{H\dot{N}}{N^{3}}+6\frac{\dot{H}}{N^{2}}+ 18
\frac{H^{2}}{N^{2}}\hspace{.08cm},\quad
A\equiv(1-3\lambda)\frac{3H^{2}}{N^{2}} \ ,
\end{equation}
and  the Hubble parameter is defined as
$H\equiv\frac{\dot{a}}{a}$.

If we now insert (\ref{cos3}) and (\ref{cos4a}) into  (\ref{cos1})
and perform variation of the action (\ref{cos1}) with respect to $N$,
we obtain
\begin{equation}\label{cos4}
\begin{split}
&\frac{3}{2}(3\lambda-1) M_{P}^2H^{2}+ M^{4}\tf+
6(3\lambda-1-\Upsilon) M^{2}H^{2}\tf'-\Upsilon M^{2}R\tf'\\
&+72\Upsilon(\Upsilon-1)H^{2}\dot{H}\tf'' +6\Upsilon^{2}
H\dot{R}\tf''=0 \ ,
\end{split}
\end{equation}
where we have set $N=1$ and $ R=6\dot{H}+12H^{2} $ \footnote{The choice
$N=1$ can be considered as the gauge fixing of the first-class
constraint $\Pi_N$ which is the generator of the scale transformation
of $N$ as follows from (\ref{PiNai}).}. Prime denotes the derivative
with respect to the argument of $\tf$, which is defined as
\begin{equation}\label{cos5}
\tf\equiv \tf\left(\frac{R_{\Upsilon}}{M^{2}} \right).
\end{equation}
The other equation, which is obtained by
 variation of the action with respect to the scale factor,
 is not an independent equation
 \footnote{The equation is the same as the
  equation which is obtained by taking time derivative of
 (\ref{cos4})
   with some algebraic manipulations.}.

We see that the structure and properties  of the (\ref{cos4}) depend
on the fact whether $\Upsilon$ is equal  to zero or not. In
particular,  if we take $\Upsilon=0$, the terms with time
derivatives vanish in  eq.(\ref{cos4}). We perform  the analysis of
this special case later and rather focus on the more standard case
when $\Upsilon\neq 0$.

\subsection{\label{sec:level1}  $\Upsilon\neq0$ case}
%As we pointed out, If we take $\Upsilon=1$ the equations must be the same as the $f(R)$-gravity.\\
%In the cosmological context, $f(R/M^{2})$-gravity theories have been used to produce accelerated expansion phase.
Since the effective equation of state parameter is defined as
\begin{equation}
w_{eff}\equiv-1-\frac{2}{3}\frac{\dot{H}}{H^{2}} \ ,
\end{equation}
there exist different mechanisms to obtain $w_{eff}<-1/3$, as follows:
\begin{itemize}
\item\textit{The de Sitter solution:} The obvious way to have
$w_{eff}<-1/3$ is to require  that a constant Hubble parameter $H_{*}$
is a solution of (\ref{cos4})
\begin{equation}\label{cos6}
\frac{3}{2}(3\lambda-1) M_{P}^2 H_{*}^{2}+ M^{4}\tf_{*}+6
(3\lambda-1-3\Upsilon) M^{2}H_{*}^{2}\tf_{*}'=0 \ ,
\end{equation}
where
\begin{equation}
\tf_{*}\equiv \tf\left(\frac{R_{\Upsilon}^{*}}{M^{2}}\right)=
\tf\left[(1-3\lambda)\frac{3H_{*}^{2}}{M^{2}}+18\frac{\Upsilon}{M^{2}}
H_{*}^{2}\right].
\end{equation}
If eq.(\ref{cos6}) has at least one solution, we should  determine
whether this solution is stable or unstable. In order to check the
stability, we consider a small perturbation $ \delta H(t)$ around the
solution as
\begin{equation}
H(t)=H_{*}+\delta H(t) .
\end{equation}
Inserting this  expression into  eq.(\ref{cos4}) and performing its
linearization, we obtain
\begin{equation}\label{cos7}
\Upsilon^{2}\delta\ddot{H}+
3H_{*}\Upsilon(\Upsilon+\lambda-1)\delta\dot{H}+ \Gamma_{\lambda}
H_{*}^{2}\delta H(t)=0 \ ,
\end{equation}
where
\begin{equation}\label{cos8}
\Gamma_{\lambda}\equiv(3\lambda-1)\left(\frac{M_{P}^{2}}{12
H_{*}^{2}\tf_{*}''}+\frac{M^{2}}{6H_{*}^{2}}\frac{\tf_{*}'}{\tf_{*}''}\right)
+(3\lambda-1-3\Upsilon)(6\Upsilon-3\lambda+1) \ ,
\end{equation}
and where we have used (\ref{cos6}). eq.(\ref{cos7}) has a solution
$\delta H\propto \exp(\chi H_{*}t$), where $\chi$ is solution of the
following equation
 \begin{equation}\label{cos9}
 \Upsilon^{2}\chi^{2}
+3\Upsilon(\Upsilon+\lambda-1)\chi+\Gamma_{\lambda}=0 \ .
\end{equation}
Thus, it is clear that the stability of the de Sitter
depends both on the specific form of the theory
and  on the values of the parameters.\\
To compare the new features of the theory with the usual $f(R)$ gravity,
 let us henceforth in this section take $\lambda=1$.
Then we see that  the second term in eq.(\ref{cos9})
  is   positive and  we have  the following possibilities:\\
For
\begin{equation}
\Gamma\equiv\Gamma_{\lambda=1}>0 \ ,
\end{equation}
the real part of the solutions or the real solutions are negative.
Thus, in this case the Sitter solution is an attractor
 solution and is suitable for the late-time cosmology. As a check
we note that for  $\Upsilon=1$, both (\ref{cos6}) and (\ref{cos8}) have the
same form as the corresponding relations derived in \cite{bf}.

For
\begin{equation}
\Gamma<0 \
\end{equation}
the equation (\ref{cos9}) has two real solutions,  where one of them,
$\chi_{+}$, is positive and the second one, $\chi_{-}$,  is negative.
Then  de Sitter solution is unstable since at late times we have
\begin{equation}
\delta H(t)\propto\exp(\chi_{+} H_{*}t) \ .
\end{equation}
In the original Starobinsky's model \cite{str}
 and as well in the context of asymptotically
 safe inflation \cite{wei},  the unstable de
 Sitter solution has been used to produce
  inflationary era for the early Universe.
   The mechanism is based on the fact that
    during the time interval $\delta t$, where
     $\chi_{+}H_{*}\delta t<1$, the solution is close to the de Sitter solution.
 Thus, one can define the number of $e-$foldings
 as $N_{e}=H_{*}\delta t$.
 In order to  solve the horizon problem,
  we take $N_{e}=60$ which gives an upper bound
   on $\chi_{+}$ and hence on $\Upsilon$ too.

 Specifically, let us consider the  following form of the function
$\tf$
\begin{equation}
\tf(R/M^{2})=-M^{2n}/R^{n} \ ,
\end{equation}
where $n$ is a positive number
 \cite{modi}. Then using (\ref{repla}), we find that the restricted version of
this theory is given by
\begin{equation}\label{cos10}
\tf=\frac{-M^{2n}}{[R+(\Upsilon-1)\Xi)]^n} ,
\end{equation}
so that from (\ref{cos6}) we obtain
\begin{equation}\label{cos11}
H_{*}^{2n+2}=\frac{M^{2n+4}}{M_{P}^{2}}\frac{12\Upsilon-6}{(18\Upsilon-6)^{n+1}}
\ .
\end{equation}
Inserting  (\ref{cos11}) into (\ref{cos8}), we find
\begin{equation}
\Gamma=-2\frac{n+1}{n}(1-3\Upsilon)^{2} \ .
\end{equation}
Therefore, for any $\Upsilon$ and $n$ the de Sitter solution is
unstable. Note also that for
 $\Upsilon=1$ our discussion is in agreement with \cite{modi}.\\
To see another example, consider $R^{2}$-gravity. In this case
(\ref{cos6}) gives
\begin{equation}\label{cos12}
H_{*}^2=\frac{M_{P}^{2}}{36(1+3\Upsilon^{2}-4\Upsilon)} \ .
\end{equation}
It is important to stress that there is no de Sitter solution in
the case when $\Upsilon=1$,  while it exists when either $\Upsilon>1$ or
$\Upsilon<1/3$, as is shown in Fig. 1.
\begin{figure}
 \includegraphics{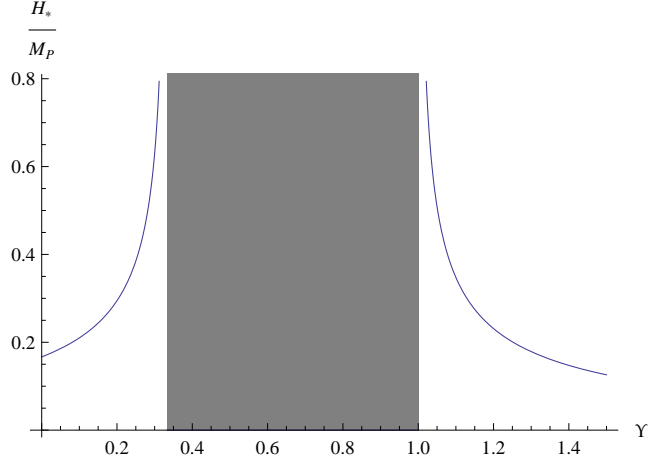}
 \centering
\caption{\label{fig:epsart} $\frac{H_{*}}{M_{P}}$ vs $\Upsilon$ for eq.(\ref{cos12}).  There is no de Sitter solution in the gray region. As is clear, the de Sitter solution can be produced by breaking the diffeomorphism symmetry.}
\end{figure}

 For such values of parameters eqs. (\ref{cos8}) and
(\ref{cos12}) give
\begin{equation}\label{cos13}
\Gamma=-3(1+3\Upsilon^2-4\Upsilon) \ .
\end{equation}
Thus, all the de Sitter solutions are unstable. Using (\ref{cos13}),
we derive the positive solution of (\ref{cos9}) in the form
\begin{equation}
\chi_{+}=-\frac{3}{2}+\sqrt{\frac{45}{4}+\frac{3}{\Upsilon^2}-\frac{12}{\Upsilon}}
\ .
\end{equation}
So,  in order to have $\chi_{+}<1/N_{e}\sim0.01$,  we have to take
$1<\Upsilon<1.1$ or $\frac{1}{3}-0.1<\Upsilon<\frac{1}{3}$ \ .
\\
\item\textit{Power-Law Acceleration:}
We have shown that for $\tf \sim R^{-n}$
 the de Sitter solution is not stable. On the other hand,
 it was shown  in \cite{modi}
  that there exists another mechanism which  produces
  the accelerated expansion phase.
We would like to investigate this mechanism in the context of
restricted the $f(R)$ gravity, while we proceed in a slightly different
from \cite{modi} way. Of course our procedure is valid for  $\Upsilon=1$
     and we  clarify  this point in \cite{note}.
 \\
By inserting (\ref{cos10}) into  (\ref{cos4}), we obtain the following
form of the equation
\begin{equation}\label{cos14}
\begin{split}
&(1+n\Upsilon)(-1+3\Upsilon+\Upsilon\frac{\dot{H}}{H^{2}})^{2}-n(2-\Upsilon)(-1+3\Upsilon+\Upsilon\frac{\dot{H}}{H^{2}})\\
&+2\Upsilon(\Upsilon-1)n(n+1)\frac{\dot{H}}{H^{2}}+n(n+1)\Upsilon^{2}(\frac{\ddot{H}}{H^3}+4\frac{\dot{H}}{H^{2}})\\
&=\frac{3M_{P}^{2}}{M^{4+2n}}H^{2+2n}(-6+18\Upsilon+6\Upsilon\frac{\dot{H}}{H^{2}})^n.
\end{split}
\end{equation}
Since there is not any stable de Sitter solution,
 as time passes the right-hand side of (\ref{cos14}) decreases.
 So, for the late-time cosmology, one can drop this term.
  But, without this term, the equation admits
a power-law solution as $a\propto t^{1/\epsilon}$, where $\epsilon
>0$ is determined from the  following equation
\begin{equation}\label{cos15}
\begin{split}
&\epsilon^2\Upsilon^2(1+n\Upsilon+2n^{2}+2n)+(3\Upsilon-1)(-1-2n+3\Upsilon+3n\Upsilon^{2})\\
&-\epsilon\Upsilon(-2
+5n\Upsilon+6\Upsilon+6n\Upsilon^2+6n^{2}\Upsilon-2n^{2}-4n)=0 \ .
\end{split}
\end{equation}
This equation has two solutions. In  one of them the denominator of
eq.(\ref{cos10}) approaches to zero and we will discuss it
afterwards. The other solution is
\begin{equation}\label{cos16}
\epsilon=\frac{3n\Upsilon^{2}+3\Upsilon-2n-1}{\Upsilon(1+2n+2n^{2}+n\Upsilon)}
\ .
\end{equation}
For the latter solution the effective equation of state parameter can be given as
\begin{equation}\label{cos17}
w_{eff}=-1+\frac{2}{3}\frac{3n\Upsilon^{2}+3\Upsilon-2n-1}{\Upsilon(1+2n+2n^{2}+n\Upsilon)}
\ .
\end{equation}
Note again that for  $\Upsilon=1$ this result agrees with the corresponding relation in  \cite{modi}. We also see that,  since   $\epsilon>0$, we can  obtain  constraints on
$\Upsilon$ from eq.(\ref{cos16}). For example if we take $n=1$,  we
obtain
\begin{equation}
\epsilon=3\frac{(\Upsilon^2+\Upsilon-1)}{\Upsilon(5+\Upsilon)} \ .
\end{equation}
Therefore,  in order to  impose $\epsilon>0$, we  should  take $\Upsilon>0.61$.\\
To show  the implication of the theory, let us compare two situations. In
the first case,  we take $\Upsilon=1$ which gives
\begin{equation}\label{cos18}
w_{eff}|_{\Upsilon=1}=-1+\frac{2}{3}\frac{n+2}{(2n^{2}+3n+1)} \ .
\end{equation}
So, for $n=1$ we have $w_{eff}|_{\Upsilon=1}=-2/3$,
which is not in agreement with the recent observations
\cite{Ade:2015lrj}. Of course, as argued in \cite{modi},
 one can increase $n$ to fit the model with the observations as is shown in Fig. 2.
  For example, if we require $w_{eff}|_{\Upsilon=1}=-0.997$
   to reconcile the model with the recent observations \cite{Ade:2015lrj},
    we should  take $n=100$,  which may  not be interesting.\\

On the other hand, let us now consider $n=1$ and leave $\Upsilon$
arbitrary. Then from (\ref{cos16}) we  obtain
\begin{equation}\label{cos19}
w_{eff}|_{n=1}=-1+2\frac{\Upsilon^{2}+\Upsilon-1}{\Upsilon(5+\Upsilon)}
\ .
\end{equation}
Here, we can change $\Upsilon$ to fit the model with the observations, as is shown in Fig. 3.
 For example, if we take $\Upsilon=0.62$,
 we have $w_{eff}|_{n=1}=-0.997$, which is in agreement with the
 recent observations \cite{Ade:2015lrj}.
\item\textit{Accelerating by $\Upsilon-\frac{1}{3}\ll1$:}
We have argued that at the late-time, $t\rightarrow\infty$, we can
neglect the right-hand side of eq.(\ref{cos15}). In addition to
(\ref{cos16}), there exists another asymptotic solution of
eq.(\ref{cos15})
\begin{equation}\label{cos20}
\epsilon\rightarrow 3-\frac{1}{\Upsilon} \  .
\end{equation}
It is important to stress that (\ref{cos16}) and
(\ref{cos20}) are the asymptotic solutions of (\ref{cos15}).
In fact, from (\ref{cos20}) we see that
  the denominator of eq.(\ref{cos10})  approaches  zero, which means
   that  the effective density of the model increases with time. From eq.(\ref{cos20}) it is clear that if we take $\Upsilon-\frac{1}{3}\ll1$, the accelerated expansion phase  emerges.\\
Let us now discuss the second asymptotic solution of  eq.(\ref{cos15})
when we take $\Upsilon=1$. In this  case,  we have
$\epsilon\rightarrow 2$, as follows from eq.(\ref{cos20}). Thus,
$w_{eff}\rightarrow1/3$, i. e. $a\rightarrow t^{1/2}$. Actually this
point for the usual $\tf(R/M^{2})$ gravity has been discussed in
\cite{tro2006}.
\end{itemize}
 \begin{figure}
 \includegraphics{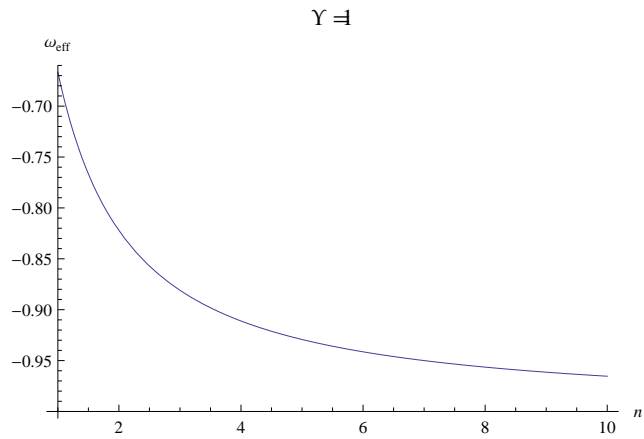}
 \centering
\caption{\label{fig:epsart} $w_{eff}|_{\Upsilon=1}$ vs $n$ for eq.(\ref{cos18}).}
\end{figure}
 \begin{figure}
 \includegraphics{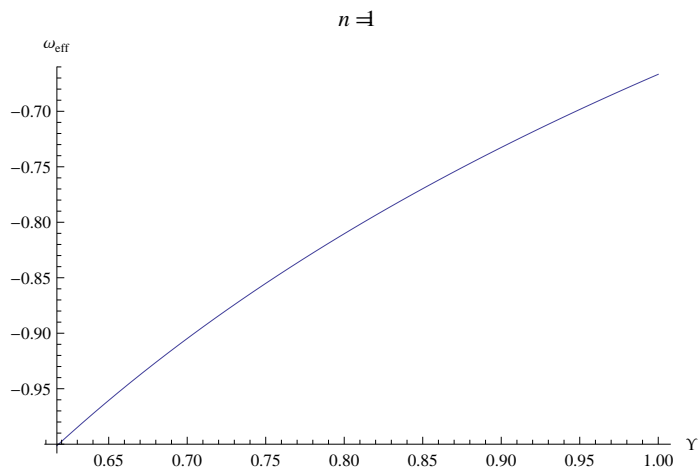}
 \centering
\caption{\label{fig:epsart} $w_{eff}|_{n=1}$ vs $\Upsilon$ for eq.(\ref{cos19}).}
\end{figure}
%55555555555555555555555555555555555555555555555555555555
%55555555555555555555555555555555555555555555555555555555
In the case of the  matter-dominated eras (radiation or the cold dark matter), it is sufficient to add the density of the matter $\rho_{M}$, to the right-hand
 side of eq.(\ref{cos4}) and consider $H=1/kt$, where $k=2$ for the radiation-dominated era and $k=3/2$ for the cold dark matter-dominated era.\\
For the specific model (\ref{cos10}), we obtain
\begin{equation}\label{mat1}
3M_{P}^2H^{2}=\rho_{M}+\rho_{eff} \ ,
\end{equation}
where
\begin{equation}\label{mat2}
\rho_{eff}\equiv\frac{M^{4+2n}}{H^{2n}(18\Upsilon-6-6\Upsilon k)^n}\times (\text{left-hand side of eq.(\ref{cos14})}) \ .
\end{equation}
Note that in the matter-dominated era
 $\frac{\rho_{eff}}{\rho_{M}}\ll 1$,
 so that for $n=1$ we obtain
\begin{equation}\label{mat3}
3M_{P}^{2}H^{2}=\rho_{M}+\Omega\frac{ M^{6}}{H^{2}}  \  ,
\end{equation}
where \cite{note1}
\begin{equation}\label{mat4}
\begin{split}
&\Omega|_{\text{radiation-dominated era}}=\frac{1}{6}(\Upsilon^{2}-7\Upsilon-3) \ ,\\
&\Omega|_{\text{dark matter-dominated era}}=\frac{1}{4}(\Upsilon^{2}-3\Upsilon-2) \ .
\end{split}
\end{equation}
Thus, using  eq.(\ref{mat3}) and $\frac{\rho_{eff}}{\rho_{M}}\ll1$, we have
\begin{equation}\label{mat5}
3M_{P}^{2}H^{2}=\rho_{M}+3\Omega\frac{M_{P}^{2}M^{6}}{\rho_{M}} \ .
\end{equation}
>From eq.(\ref{mat3}) or eq.(\ref{mat5}), it follows that in the
matter-dominated eras, $\rho_{eff}\rho_{M}\propto1$. So, eventually
$\rho_{eff}$ will be dominated and the mechanism for  the power-law
acceleration can occur.
\subsection{\label{sec:level1} $\Upsilon=0$ case}
 Let us now focus our attention on the special case $\Upsilon=0$.
 This special case was previously studied in \cite{Gao} in a different from ours approach .
%Here we present a nontrivial extension of this model which connects this case to the Chaplygin gas \cite{chap}.\\

To begin with, we  note that eq.\eqref{cos4} is valid for any  $\Upsilon$. On the
other hand,  for $\Upsilon=0$ this equation
 reduces  to an algebraic equation for the Hubble parameter. So, if the equation has a solution we find that it is the de Sitter solution which is suitable for the late-time cosmology.\\
 For instance, consider \eqref{cos10} for $n=1$ and $\Upsilon=0$. Then equation \eqref{cos4} with the  matter density on the right-hand side yields
\begin{equation}\label{op1}
3M_{P}^{2}H^{2}=\rho_{M}+\frac{1}{6}\frac{ M^{6}}{H^{2}} \ .
\end{equation}
This  equation is similar to eq.\eqref{mat3}, but note that  the equation
(\ref{op1})
 is valid during all the cosmological eras. Solving this equation for $H^2$, we find
\begin{equation}
6M_{P}^{2}H^{2}=\rho_{M}+\sqrt{\rho_{M}^{2}+2 M_{P}^{2} M^{6}}.
\end{equation}
So, at the late-time we have
\begin{equation}\label{1-6}
3H^{2}\rightarrow\sqrt{\frac{1}{2}}\frac{ M^{3}}{M_{P}} \ .
\end{equation}

\section{Discussion}\label{sixth}
Let us outline the main results of the paper. We have analyzed the
recently proposed version of $f(R)$ gravity with broken
four-dimensional diffeomorphism  by changing  the constant in front
of the total divergence term that arises in the
($3+1$)-decomposition of scalar curvature. We have shown that this
naive modification of theory is not consistent from the Hamiltonian
point of view due to the fact that the Hamiltonian constraint is a
second-class constraint with itself. We have proposed two ways how
to resolve this issue. The first one is based on the observation
that the Poisson bracket between the Hamiltonian constraint vanishes
when we impose an additional constraint on the scalar field $B$.
However, a  careful Hamiltonian analysis shows that the restricted
$f(R)$ gravity with this additional constraint is equivalent to the
ordinary Einstein-Hilbert action. Further, we have argued that the
right way how to correctly define the restricted $f(R)$ theory of
gravity is to include the terms which are invariant under the
spatial diffeomorphism , for example, the gradient of the lapse. We
have performed the Hamiltonian analysis of this  theory and we have
found that it is consistent one  from the Hamiltonian point of view
and  have shown that this theory is equivalent to the low-energy
limit of non-projectable $f(R)$ HL gravity. Moreover, we have
identified two global first-class constraints, which ensure that the
Hamiltonian is invariant under the global time reparametrization and
global rescaling of lapse. Finally, we have  discussed some
cosmological application of the restricted $f(R)$ gravity  and have
found several  interesting implications. In particular, we have
discussed the differences between the usual $R^{n}$ gravity, with $
n < 0 $  and its corresponding restricted version. In addition, it
has been shown  how the asymptotic solutions of $R^{n}$ gravity can
be changed by the broken symmetry. Moreover, we have  found that it
is possible to find the de Sitter solution in the case of $R^2$
gravity,  which does not exist in the case of $\Upsilon=1$. It has
been also found that with a suitable choice of the parameter
$\Upsilon$, this solution describes the inflation phase of cosmology
with the correct number of e-foldings. These results imply nice
physical meaning of the parameter $\Upsilon$: For the early Universe
$\Upsilon$ determines the scale of inflation and the stability of de
Sitter solution (see eq.(\ref{cos12})), while for the late-time
cosmology, $\Upsilon$ determines the parameter in  the equation of
state, which can be measured (see eq.(\ref{cos17}) and discussion
after it).

The results presented are  encouraging and  the cosmological
applications in the context of restricted $f(R)$ gravity certainly desrve to be
studied further and in more details. Specifically, one should
analyze the fluctuations around the cosmological solutions in this
theory. We  expect that there is an additional scalar mode and its behaviour
should be analyzed. It would also be interesting  to analyze this
mode around the flat background following the corresponding analysis in
the  case of the healthy extension of  non-projectable HL gravity
\cite{Blas:2010hb}. We hope to return to this problem in future.

 \noindent {\bf Acknowledgements}
\\
The work of J.K. was supported by the Grant Agency of the Czech Republic
under the grant P201/12/G028.
%%%%%%%%%%%%%%%%%%%%%%%%%%%%%%%%%%%%%%
%%%%%%% Thebibligraphy %%%%%%%%%%%
%%%%%%%%%%
%%%%%%%%%%%%%%%%%%%%%%%%%%%%%%%%%%%%%


\begin{thebibliography}{20 }
\bibitem{note}
In Ref. \cite{modi}, the authors study $f(R)=-M^{2n}/R^{n}$ by using the so-called the Einstein-frame.
 Here, we do not use this frame. For example from eq. \eqref{cos14} to eq. \eqref{cos15}, we use an approximation. To find what is the meaning of the similar approximation when we take $\Upsilon=1$ in the Einstein-frame, see discussion of Ref. \cite{modi} that  the authors argued that "\textit{..Soon thereafter, the potential is well-approximated by...}".
\bibitem{tro2006}
L.~Amendola, D.~Polarski, S.~Tsujikawa,\emph{``Are f(R) Dark Energy
Models Cosmologically Viable?,''} Phys.\ Rev.\ Lett. \textbf{98},
131302, (2007). Note that the authors of this paper considered the
usual $f(R/M^{2})=\frac{M^{2n}}{R^{n}}$ gravity with the matters.
They showed that at the late-time, when the density of the matters
are diluted by the expansion of the universe, in addition to eq.
\eqref{cos18}, it is possible that $w_{eff}\rightarrow 1/3$. Thus it
is unclear that we have $w_{eff}<-1/3$ after the dark
matter-dominated era in the such models. Note that for this case, in
this reference, as $w_{eff}\rightarrow 1/3$ we have $R\rightarrow0$.
\bibitem{note1}
To find $\Omega|_{\text{radiation-dominated era}}$ in eq.
\eqref{mat4}, we have a factor as $\frac{1-\Upsilon}{1-\Upsilon}$.
For this reason, we cannot obtain this term if we take $\Upsilon=1$.
Actually, $f(R)=-M^{2n}/R^{n}$ has this problem in the
radiation-dominated ere, because in this era $R=0$. One can regard
this note as an advantage of our model.
\end{thebibliography}
\end{document}